# Quasi-1D Modeling of Polymer Melt Die Swell in Short Dies

JAE-HYEUK JEONG *and* ARKADY I. LEONOV[†]

*Department of Polymer Engineering*
*The University of Akron*
*Akron, Ohio 44325-0301*

**Abstract**

This paper describes the isothermal die swell using our recent quasi-1D model for fast (high Deborah number) contraction flows of polymers melts. Because the model analyzes the flow in several flow regions as one continuous process, it makes possible to evaluate the die swell as a qusi-1D extrudate flow in dies of various lengths. Using the asymptotic matching condition for the change in flow type at the die exit allowed us to find the swelling profile for extrudate along the flow direction. The calculations in paper performed using a multi-mode viscoelastic constitutive equation of differential type, are compared with the experimental/direct numerical data including basic rheological tests. The presented swelling model involves no fitting parameter and is applicable for calculations using any viscoelastic constitutive equation.

## 1. Introduction

The die swell (or extrudate swell, or Barrus Effect) occurs when a polymer liquid, discharged from die/channel, experiences a free surface flow. The swell is commonly observed in many polymer processing operations, such as capillary/slit type process, fiber spinning, extrusion, etc. It originates from elastic nature of polymer fluids and is qualitatively explained as the release of liquid elastic energy stored in die flow, after a rapid restructuring the flow type at the die exit. Usually the swell is characterized by the measurable parameter *B* presented the ratio of cross sectional areas of extrudate to the die.

It is well known that the swell might be derived from known (nonlinear) deformation history of polymeric liquids under certain geometrical/operation conditions. The knowledge of this history is especially important when the flow Deborah number (*De*) and the contraction ratio in contraction flow are very high, and the die is relatively short. In this case, the flow in the die is not developed and extrudate keeps the memory of early deformation at the entrance region. Therefore the common assumption of the polymer flow as developed near the die exit, employed in many theoretical/computational papers, is invalid and swelling has to be analyzed using continuous computations of reservoir/channel viscoelastic flows.

[†] Corresponding author.
*E-mail addresses*: jeong@uakron.edu, leonov@uakron.edu

Various theoretical approaches have been developed for evaluation of the swell ratio. Early Bagley *et al* [1] assumed that the swelling results from the geometry-influenced "memory effect". They empirically expressed the equilibrium swell ratio via shear strain and die geometry. Later Bagley and Duffey [2] improved the previous approach [1] including into consideration the Mooney energy function. Similar to Lodge [3], Tanner [4] treated the swelling as a process of recovering elastic energy after removing shearing. Using force balance in the swollen state and K-BKZ constitutive equation, he expressed the equilibrium swelling ratio as a function of the first normal stress difference and shear strain ratio for polymer flow in a tube of infinite length. The text [5] and paper [6] involved the Tanner approach in their empirical formulations to analyze geometrical effects.

More recently, Garcia-Region *et al* [7,8] measured swelling ratio depending on distance/time from exiting die. Leonov and Prokunin [9] measured this ratio for a PIB and analyzed it theoretically, using the assumption of developed flow in the die. Using the force balance and equalizing the elastic energy flux between the die shear flow and the extrudate free jet, they calculated the increase in swell with growing distance from the die. Their theoretical predictions overestimated, however, the experimental data. Choi *et al* [10] discussed the stability and flexibility of their method of swell calculations, employing Leonov multi-mode constitutive equation and using the Tanner method [4] for evaluating the instant component of swell for steady viscoelastic flow in a die. Nevertheless, the calculations of [10] overestimated the experimental data [9]. A possible reason for those disagreements is incomplete material characterization. We will revisit the problem below to improve the results of calculations using our scheme with more precise evaluation of material parameters.

Computational stability was a big problem at early stage of direct computations. Crochet and Keunings [11] demonstrated the instability of a numerical algorithm in solving the die swell problem for large *De* number flow. Using upper convected Maxwell model, they could obtain the results only when $De \lesssim 1$. Later they increased their computational capability using Oldroyd B [12] and KBKZ [13] constitutive models. Bush *et al* [14] related the computational instability at high *De* number extrudate flows to the formulation of viscoelastic constitutive equations (CE's) and concluded that the reason of failure of direct computations came not only from the flaw of numerical algorithm but also from the instability of used CE's. They found that MPTT and modified Leonov models are capable to describe high *De* number flows while agreeing well with experimental data. Also Wesson *et al* [15] independently tested two different types of CE's in their numerical study concerning the instability of CE's. All above findings have been recently confirmed within the general theory of CE stability [16].

Another problem of direct numerical analysis of swell in flows of viscoelastic liquids is exhaustive computing time. To avoid this, an alternative numerical procedures were elaborated that tried using stream function [17,18,19,20 and 21]. Some of the authors [18] pointed out that the extrudate flow is not the major computational problem but the restructuring flow in the die is. Since they considered the contraction flow as a complex flow with interaction between the shear and extension, they select the Wagner integral type model as a proper CE that has been widely tested in dynamic shearing and extension experiments. Employing Protean coordinate system and uncertain damping function with fitting parameters at the tapered transient entrance region, they extended the computational results for the extrudate swell and melt spinning for *De* number over 1000.

Recently the importance of extension in the die entrance flow and swelling has been recognized in many publications and handled in various ways. However, for the best of our knowledge, just few attempts have been made treating the whole flow region as one continuous process. In this paper we extend our contraction flow model [22] for the high *De* number flows in short dies. We do not assume here the existence of developed flow in the die and make modeling predictions of swell as continuous calculations of fast contraction flow of polymer melts in the whole flow region. In addition, we will test the descriptive ability of chosen stable CE on existing basic experimental data. A remarkable feature of our 1D contraction/swelling model is that it describes high *De* number flows of polymers without using any adjustable parameter.

## 2. Constitutive Equation

We employ in this paper a multi-mode viscoelastic CE of differential type which is briefly discussed below (for more details see Refs.[23-25]). This CE has been used in our paper [22] for simplified calculations of contraction flow.

For each relaxation mode with relaxation time $\theta_k$ and elastic modulus $G_k$, the CE operates with a modal elastic Finger tensor $\underline{\underline{c}}_k$ whose evolution in incompressible case is:

$$2\theta_k \stackrel{\nabla}{\underline{\underline{c}}}_k + b[\underline{\underline{c}}_k^2 + \underline{\underline{c}}_k(I_{2k} - I_{1k})/3 - \underline{\underline{\delta}}] = \underline{\underline{0}}, \quad \stackrel{\nabla}{\underline{\underline{c}}} \equiv \dot{\underline{\underline{c}}} - \underline{\underline{c}} \cdot \nabla v - (\nabla v)^T \cdot \underline{\underline{c}}$$

$$(I_{1k} = tr\underline{\underline{c}}_k, \quad I_{2k} = tr\underline{\underline{c}}_k^{-1}, \quad I_{3k} = \det \underline{\underline{c}}_k = 1; \quad b = b(I_1, I_2)) \tag{1}$$

Here $\nabla v$ is the velocity gradient tensor, $\underline{\underline{\delta}}$ is the unit tensor, and *b* is a mode independent positive scaling factor for the relaxation time, which is generally a function of invariants, $I_{1k}$ and $I_{2k}$.

The extra stress tensor $\underline{\underline{\sigma}}_e$ and the total stress $\underline{\underline{\sigma}}$ are defined as:

$$\underline{\underline{\sigma}} = -p\underline{\underline{\delta}} + \underline{\underline{\sigma}}_e; \quad \underline{\underline{\sigma}}_e = \sum_k \underline{\underline{\sigma}}_k = 2\sum_k G_k \underline{\underline{c}}_k \cdot \partial w_k / \partial \underline{\underline{c}}_k.$$

$$W \equiv G(T)w(I_1) = \frac{3G(T)}{2(n+1)}[(I_1/3)^{n+1} - 1] \tag{2}$$

Here *p* is the isotropic pressure and $\underline{\underline{\sigma}}_k$ is the extra stress tensor for each relaxation mode. The elastic potential, *W*, for each $k^{th}$ nonlinear relaxation mode is represented via the Hookean modulus, $G(T)$, and the non-dimensional function, *w*, which is characterized by mode-independent numerical parameter *n*. In the cases considered below, the non-dimensional values, *b* and *n*, are used consistently as shown in Table 1 to describe the rheological data. Then the only the linear relaxation spectrum, $\{G_k, \theta_k\}$ should be found for complete rheological characterization of a polymer melt. Interested readers may find more information of the above CE in Refs.[23-24].

## 2. Preliminary Analysis on Channel Flow: Quasi-1D Model of Contraction Flow

The entire contraction flow is analyzed below using our recent quasi-1D model [22]. We now briefly discuss this model.

When the reservoir is long enough, polymer melts demonstrate a developed steady shearing ("*far field*") entrance flow in the reservoir pre-entrance region (i). Additionally, there is a reservoir entrance region (ii) near the die entrance where the main stream converges to the die ("*near field entrance flow*). This flow sometimes is accompanied by weak entrance vortices at the flow periphery near the entrance corner. When $x_1$ be the axial coordinate directed downstream along the symmetry axis with the origin ($x_1 = 0$) at the die orifice, the far and near field flows are located in the respective axial domains ($x_1 < -l$) for the region (i) and ($0 > x_1 > -l$) for the region (ii); the coordinate, $x_1 = -l$, being the parameter searched for. Although the change of the flow characters from the far field to the near field entrance flow needs a small transitional axial length $\delta$ (<<$l$), we asymptotically use in this model the value $x_1 = -l$ as the boundary where sudden change in flow type happens.

The contraction flow in the near field entrance region (ii) is initially analyzed as an inhomogeneous extensional (or "jet") flow that commonly occurs either in polymer fiber spinning or in polymer sheet processing operations. It is assumed that such an approach is approximately valid when *De* number of flow is high enough. However, for relatively low *De* number flows, the simple jet model has been complemented with analysis of secondary circulatory flows in the corners, which produce additional drag acting on primary jet.

Two types of reservoir geometries are analyzed in Ref.[22] and in this paper: the slit type reservoir (Fig.1a) with cross-sectional sizes $2L_R$ (thickness) and $L_W$ (width) where $L_R << L_W$, and the reservoir of circular geometry with the reservoir radius $R_R$. To avoid unnecessary duplications and simplify notations, a brief description of simultaneous analysis of flows in both types of geometries [22] is presented below, with asterisked formulae attributed to the circular geometry. The same notations $\{x_1, x_2, x_3\}$ are used below for the Cartesian and cylindrical coordinate systems to describe the respective flows in the slit ($-L_R < x_2 < L_R$) and circular ($0 < x_2 < R_R$) geometries. The index 1 stands for axial ($z$), 2 – for transversal ($y$ or $r^*$), and 3 – for neutral ($x$ or $\varphi^*$) directions.

## 2.1. Flow in the Far Field Entrance Region (i):

$$\{x_1 < -l,\ L_R > x_2 > -L_R\ \text{or}\ 0 < x_2 < R_R\}.$$

The reservoir flow in the region (i) is the steady shear flow. The known solution of CE (1) for steady flow is:

$$c_{11,k} = \frac{\sqrt{2}\zeta_k}{\sqrt{\zeta_k+1}},\ c_{22,k} = \frac{\sqrt{2}}{\sqrt{\zeta_k+1}},\ c_{12,k} = \frac{\sqrt{\zeta_k^2-1}}{\zeta_k+1},\ c_{33,k} = 1,\ \zeta_k(\dot\gamma) = \sqrt{1+(2\theta_k\dot\gamma/b)^2} \quad (3)$$

$$\sigma_{ij,e} = \sum_k G_k(I_{1,k}/3)^n c_{ij,k};\quad \dot\gamma = dv_1/dx_2;\ I_{1,k} = I_{2,k} = c_{11,k} + c_{22,k} + 1 = \sqrt{2(\zeta_k+1)} + 1.$$

Here $\underline{\nabla v}$ is velocity gradient, $\underline{\underline{\sigma}}_e$ the extra stress, $\underline{\underline{c}}_k$ the elastic strain, $\dot\gamma$ the shear rate and $v_1$ is the only non-zero velocity component.

The momentum balance equations result in:

$$\sigma_{12,e}(x_2) = \frac{dP}{dx_1}\cdot x_2;\quad \sigma_{12,e}(x_2)^* = \frac{1}{2}\frac{dP}{dx_1}\cdot x_2 \tag{4,4*}$$

Here $P$ is the pressure and $\sigma_{12,e}$ is the shear stress. The expression for flow rate $Q$ is:

$$Q = 2L_W\int_0^{L_R} v_1 dx_2 = -2L_W\int_0^{L_R} x_2\dot\gamma dx_2;\quad Q^* = 2\pi\int_0^{R_R} v_1 x_2 dx_2 = -\pi\int_0^{R_R} x_2^2\dot\gamma dx_2. \tag{5,5*}$$

Here the constant $Q$ is considered below as given. When $Q$ is specified, the shear rate, $\dot\gamma(x_2)$, can be found from Eqs.(3)-(5). It enables us to compute the complete profiles of rheological variables for the steady shear flow up to the distance $l$ from the entrance.

## 2.2. Flow in the Near Field Entrance Region (ii)

$$\{0 > x_1 > -l,\ L_R > x_2 > -L_R\ (0 < x_2 < R_R)\}.$$

### 2.2.1. *Jet Flow Model*

In the slit die, the jet flow is presented as an inhomogeneous planar elongation flow, and in the circular die as an inhomogeneous simple elongation. Then, in both cases the modal evolution equations are expressed using convective approximation for time derivative and the incompressibility condition in Eq. (1) as:

$$\frac{Q}{A\lambda_k} \cdot \frac{d\lambda_k}{dx_1} + \frac{b}{4\theta_k}(\lambda_k^2 - \frac{1}{\lambda_k^2}) = -\frac{Q}{A^2} \cdot \frac{dA}{dx_1};  \qquad (6)$$

$$\frac{Q}{A\lambda_k} \cdot \frac{d\lambda_k}{dx_1} + \frac{b}{6\theta_k}(\lambda_k^2 - \frac{1}{\lambda_k})(1 + \frac{1}{\lambda_k}) = -\frac{Q}{A^2} \cdot \frac{dA}{dx_1}. \qquad (6^*)$$

Here $\lambda_k$ is the elastic strain, $\underline{\nabla v}$ the velocity gradient tensor, $\underline{\underline{\sigma}}_e$ the extra stress tensor, $\dot{\varepsilon} = dA/Adx_1$ is the elongation rate, $Q$ is the flow rate, and $A$ is the jet cross-sectional area. The expression for the elongation stress $\sigma_{ext}$, in the multi-mode case used in calculations is presented as:

$$\sigma_{ext} = \sum_k G(I_{1,k}/3)^n \cdot (\lambda_k^2 - \lambda_k^{-2}), \qquad I_{1k} = 1 + \lambda_k^2 + \lambda_k^{-2}; \qquad (7)$$

$$\sigma_{ext}^* = \sum_k G(I_{1,k}/3)^n \cdot (\lambda_k^2 - \lambda_k^{-1}), \qquad I_{1k}^* = \lambda_k^2 + 2\lambda_k^{-1}. \qquad (7^*)$$

Similarly to the case of inhomogeneous elongation flow, the elongation force acting along the jet is defined as, $F_{ext} = \sigma_{ext} \cdot A$. Then

$$F_{ext} \equiv A \cdot \sigma_{ext} = A \cdot \sum_k G(I_{1,k}/3)^n \cdot (\lambda_k^2 - \lambda_k^{-2}); \quad I_{1k} = 1 + \lambda_k^2 + \lambda_k^{-2} \qquad (8)$$

$$F_{ext}^* \equiv A^* \cdot \sigma_{ext}^* = A \cdot \sum_k G_k(I_{1k}/3)^n \cdot (\lambda_k^2 - 1/\lambda_k); \quad I_{1k}^* = \lambda_k^2 + 2/\lambda_k. \qquad (8^*)$$

The assumption $F_{ext} = const$ is related to the "jet approach" in Ref [22]. In this case,

$$F_{ext} \equiv A \cdot \sigma_{ext} = const \qquad (9)$$

When the effect of drag force on the jet flow is taken into account, the local force balance is [22]:

$$dF_{ext} = d(\sigma_{ext} \cdot 2L_{Ex}L_W) = -2\sigma_{12,surf}(x_1) \cdot L_W dx_1, \qquad (10)$$

$$dF_{ext}^* = d(\sigma_{ext}^* \cdot \pi R_{Ex}^2) = -2\pi\sigma_{12,surf}^*(x_1) \cdot R_{Ex} dx_1 \qquad (10^*)$$

Here $\sigma_{ext}$ is the extensional stress acting on the jet, $\sigma_{12,surf}$ is the shear stress at the jet surface due to circulatory drag flow, $L_W$ is the width of the slit die and $L_{Ex}$ ($R_{Ex}$) is the half thickness (or the radius) of jet flow.

Appropriate boundary conditions for these equations are:

$$x_1 = -l: \quad A = A_l = 2L_W \cdot L_R, \quad \lambda_k = \lambda_k^l; \quad A = A_l = \pi R_R^2, \quad \lambda_k = \lambda_k^l. \quad (11,11^*)$$

$$x_1 = 0: \quad A = A_0 = 2L_W \cdot L_D, \quad \lambda_k = \lambda_k^o; \quad A = A_0 = \pi R_D^2, \quad \lambda_k = \lambda_k^0 \quad (12,12^*)$$

where $L_R$ ($R_R$) is the half thickness (radius) of reservoir and $L_D$ ($R_D$) is the half thickness (radius) of die.

The set of Eqs (6) and (9) (or (10)) represent a closed system with boundary conditions described by Eq. (11) and (12).

### 2.2.2. *Matching condition at the boundary between the two entrance flow regions*

The jet flow model has not yet been completed because the initial elastic stretches $\lambda_k^l$ ($k = 1, 2, ...$) in Eq.(11) are unknown. To determine them we employ an energetic matching condition [22] at the unknown boundary, $x_1 = -l$.

$$x_1 = -l: \quad \frac{1}{A_l} \int_\Omega v_1^{sh} \cdot W_k^{sh} dA = v_1^{jet} \cdot W_k^{jet}. \quad (13)$$

Here $W_k^{sh}$ and $W_k^{jet}$ are elastic potentials for each $k^{th}$ nonlinear Maxwell mode in the far-field shearing and near-field jet flows, $v_1^{sh}$ and $v_1^{jet}$ are the axial velocities of shear and jet flow, and $A_l$ is the reservoir cross-sectional area. Using formulae (3) and (9) for simple shearing and elongation flows, reduce Eq.(13) for any $k^{th}$ relaxation mode to the form:

$$x_1 = -l: \quad \frac{L_W}{Q} \int_0^{L_R} v_1^{sh} (c_{11,k} + c_{22,k} + 1)^{n+1} dx_2 = [(\lambda_k^l)^2 + (\lambda_k^l)^{-2} + 1]^{n+1}, \quad (k = 1, 2, ...); \quad (14)$$

$$x_1 = -l: \quad \frac{2\pi}{Q} \int_0^{R_R} x_2 \cdot v_1^{sh} (c_{11,k} + c_{22,k} + 1)^{n+1} dx_2 = [(\lambda_k^l)^2 + 2(\lambda_k^l)^{-1}]^{n+1}, \quad (k = 1, 2, ...). \quad (14^*)$$

Here we used the fact that at $x_1 = -l$, the jet cross sectional area coincides with that for the reservoir. Equation (14) allows finding the initial jet elastic stretches $\lambda_k^l$ from the known elastic strain tensor profile $c_{ij,k}$ for the far field shear flow has been described by Eqs. (3)-(5). Other matching conditions are applied for changes in the flow type at the die entrance and exit.

### 2.3. Modeling developing flow in the die: { $L > x_1 > 0$, $L_D > x_2 > -L_D$ ($0 < x_2 < R_D$)) }

We model the developing flow in the die for both the slit and circular geometries as a version of non-steady viscoelastic Poiseuille flow, where the time derivative is substituted by the convective space derivative: $d/dt \approx \overline{v}_1 \partial/\partial x_1$. Here $\overline{v}_1 = Q/A_0$ is the die average longitudinal velocity, and $A_0$ is the cross-sectional die area equal to $2L_W \cdot L_D$ or $\pi(R_D)^2$ for the slit and circular die geometries, respectively. Then the *evolution equation* for each $k$th mode (the index $k$ is omitted) for both the die geometries, with the structure of elastic strain tensors $\underline{\underline{c}}_k$ shown in Eq.(1), has the form:

$$\begin{cases} \overline{v}_1 \dfrac{\partial c_{12}}{\partial x_1} + \dfrac{b}{2\theta} c_{12}(c_{11}+c_{22}) = c_{22} \dfrac{\partial v_1}{\partial x_2} & \text{("12" component)} \\ \overline{v}_1 \dfrac{\partial c_{22}}{\partial x_1} + \dfrac{b}{2\theta}(c_{12}^2 + c_{22}^2 - 1) = 0 & \text{("22" component)} \\ c_{11}c_{22} - c_{12}^2 = 1 & \text{incompressibility condition} \end{cases} \qquad (15)$$

Due to Eq. (2), the shear and longitudinal normal stress components for the extra stress tensor are:

$$\sigma_{12} = \sum_k G_k (I/3)^n c_{12,k}; \quad \sigma_{11} = \sum_k G_k (I/3)^n c_{11,k}. \qquad (16)$$

As the consequence of (15) and (16), we further use only the longitudinal component of the momentum balance equation,

$$\frac{\partial p}{\partial x_1} = \frac{\partial \sigma_{11}}{\partial x_1} + \frac{\partial \sigma_{12}}{\partial x_2}; \quad \frac{\partial p}{\partial x_1} = \frac{\partial \sigma_{11}}{\partial x_1} + \frac{1}{x_2}\frac{\partial(x_2\sigma_{12})}{\partial x_2} \qquad (17,17^*)$$

and the continuity equation:

$$\frac{\partial v_1}{\partial x_1} + \frac{\partial v_2}{\partial x_2} = 0; \quad \frac{\partial v_1}{\partial x_1} + \frac{1}{x_2}\frac{\partial(x_2 v_2)}{\partial x_2} = 0 \qquad (18,18^*)$$

Eqs.(18,18*) result in:

$$Q = 2L_W \int_0^{L_D} v_1 dx_2 = -2L_W \int_0^{L_D} x_2 \dot{\gamma} dx_2 = const; \qquad (19)$$

$$Q = 2\pi \int_0^{R_D} x_2 v_1 dx_2 = -\pi \int_0^{R_D} x_2^2 \dot{\gamma} dx_2 = const. \qquad (20^*)$$

The non-slip boundary conditions for the components of velocity are:

$$x_2 = \pm L_D: \quad v_1 = v_2 = 0,; \quad x_2 = R_D: \quad v_1 = v_2 = 0. \qquad (21,21^*)$$

In order to find the boundary conditions at the die entrance $x_1 = 0$ we use once again the matching condition for the flow type change where the jet flow turns into the shearing flow. These conditions are [22]:

$$x_1 = 0: \quad v_1^0 = \bar{v}_1 \quad (v_2^0 = 0); \quad c_{12}^0 = 0, \quad c_{33}^0 = 1, \quad c_{11}^0 = 1/c_{22}^0 = (\lambda^0)^2 \tag{22}$$

$$x_1 = 0: \quad v_1^0 = \bar{v}_1 \quad (v_2^0 = 0); \quad c_{12}^0 = 0, \quad c_{33}^0 = 1, \quad c_{11}^0 = 1/c_{22}^0 = J(\lambda^0) + \sqrt{J^2(\lambda^0) - 1};$$
$$J(\lambda^0) = 1/2[(\lambda^0)^2 + 2/\lambda^0]. \tag{22*}$$

Here $\lambda_k^0$ are the components of the elastic stretching tensor for each mode at $x_1 = 0$ from the near field entrance jet flow, $\bar{v}_1$ is the rate average velocity in the die, and all the zero superscripts denote the variables at the entrance.

## 3. Modeling of Isothermal Die Swelling Flow

As discussed earlier, the elastic energy stored in the die flow of polymer melts depends on viscoelastic characteristics of polymer melts and complete flow conditions. Therefore the swelling after flow in short dies has to be analyzed as continuous process that takes into account all the aspects of previous flow. The modal elastic Finger tensors for the channel flow of a viscoelastic liquid have been determined from our recent work on 1D quasi-state model for isothermal contraction and die flow [22]. So we will treat the modal elastic Finger tensors $\underline{\underline{c}}_k$ at die exit as known parameters.

When a polymer melt exits the die, the shear flow of the polymer in the die changes for the extrudate flow of a inhomogeneously uniaxial elongation type. Because the restructuring zone at the end of the die is small enough, this flow restructuring is approximated as in [22] as a sudden change that keeps constant the energy flux, averaged over cross-section of the die. Because of similarity of this matching condition to that described by Eq.(13) that has been discussed before for 1D quasi-steady model for contraction flow, we would hold the same matching condition here. Nevertheless, we should also take into account the instant component of free swell exiting the die, which did not happen in the situation described by Eq.(13). In order to manage this complicated situation, we introduce a two-step approach to transformation of die shearing flow into extrudate flow: (i) the transformation of shear to the elongation flow at the very end of the die, $x_1 = L - 0 \equiv L^-$, and (ii) instant strain recovery at the die outlet, $x_1 = L + 0 \equiv L^+$. Once again, we simplify the time derivative to the convective derivative in our quasi-1D model.

**3.1. Extrudate swelling:** $\{x_1 > L\}$

Unlike planar flow type in the slit entrance region, the flow extruded from slit channel is a type of uniaxial extensional flow. Therefore the same formulae for elastic strain tensors $\underline{\underline{c}}_k^{sh}$, velocity gradient tensor $\underline{\nabla v}$, and the extra stress tensor $\underline{\underline{\sigma}}_e$ are valid for both cases of swelling after polymer flow in circular or slit die. These formulae are:

$$\underline{\underline{c}}_k = \begin{pmatrix} \lambda_k^2 & 0 & 0 \\ 0 & \lambda_k^{-1} & 0 \\ 0 & 0 & \lambda_k^{-1} \end{pmatrix}, \quad \underline{\nabla v} = \dot{\varepsilon}\begin{pmatrix} 1 & 0 & 0 \\ 0 & -1/2 & 0 \\ 0 & 0 & -1/2 \end{pmatrix}, \quad \underline{\underline{\sigma}}_e = \begin{pmatrix} \sigma_{11} & 0 & 0 \\ 0 & \sigma_{22} & 0 \\ 0 & 0 & \sigma_{33} \end{pmatrix} \quad (23)$$

Here $\dot{\varepsilon}$ is the elongation rate. Inserting Eq.(23) into Eq.(1) and employing the "convective approximation", $d/dt \approx \overline{v}_1 d/dx_1$, yields:

$$\frac{1}{\lambda_k} \cdot \overline{v}_1 \frac{d\lambda_k}{dx_1} + \frac{b}{6\theta}(\lambda_k^2 - \frac{1}{\lambda_k})(1 + \frac{1}{\lambda_k}) = \dot{\varepsilon}. \quad (24)$$

Here $\overline{v}_1$ is the axial velocity averaged over the jet cross-section, and $\dot{\varepsilon} = d\overline{v}_1/dx_1$ is the elongation rate. The expression for the elongation stress (a total axial stress in jet cross-sections), $\sigma_{ext}$, in the multi-mode case is:

$$\sigma_{ext} = \sum_k G_k (I_{1,k}/3)^n \cdot (\lambda_k^2 - \lambda_k^{-1}), \quad I_{1k} = \lambda_k^2 + 2\lambda_k^{-1}. \quad (25)$$

Here $G_k(T)$ is the Hookean modulus, and $I_{1k}$ the first invariant of tensor $\underline{\underline{c}}_k$ for $k^{th}$ mode. The incompressibility condition yields:

$$\overline{v}_1(x_1) = Q/A(x_1), \quad \dot{\varepsilon} \equiv \frac{d\overline{v}_1}{dx_1} = -\frac{Q}{A^2} \cdot \frac{dA}{dx_1}. \quad (26)$$

Here $A(x_1)$ is the cross-sectional area of extrudate flow and $Q$ is given (constant) flow rate. With Eq.(26), the evolution equations (24) for each relaxation modes take the form:

$$\frac{Q}{A\lambda_k} \cdot \frac{d\lambda_k}{dx_1} + \frac{b}{6\theta}(\lambda_k^2 - \frac{1}{\lambda_k})(1 + \frac{1}{\lambda_k}) = -\frac{Q}{A^2} \cdot \frac{dA}{dx_1}. \quad (27)$$

When a constant force $F = F_0$ is applied to the free swelling extrudate,

$$F_0 \equiv A \cdot \sigma_{ext} = A \cdot \sum_k G_k (I_{1k}/3)^n \cdot (\lambda_k^2 - 1/\lambda_k); \quad I_{1k} = \lambda_k^2 + 2/\lambda_k. \tag{28}$$

If the extrudate swells free, $F_0 = 0$ in (28).

Eqs.(27) and (28) present a complete set for the unknown variables, $\lambda_k$ and $A$. The initial conditions for this problem are:

$$x_1 = L^+: \quad A = A^+, \quad \lambda_k = \lambda_k^+ \quad (k = 1, 2, ...), \tag{29}$$

where the initial conditions $A^+$ and $\lambda_k^+$ should yet to be determined.

It is also convenient to use the formulae, $\dfrac{\dot{A}}{2A} = \dfrac{\dot{L}_2}{L_2} = \dfrac{\dot{L}_3}{L_3}$ or $\dfrac{\dot{A}}{2A} = \dfrac{\dot{R}}{R}$, that follow from incompressibility conditions for time dependent simple elongation for both the geometries. Here $2L_2$ and $L_3$ are the lengths of extrudate in transversal and neutral directions for the polymer extruded from the slit die, and R is the extrudate radius for cylindrical extrudate.

### 3.2. Matching condition at the boundary between the die flow and the non-recovered extrudate: $\{x_1 = L^-, \ L_D > x_2 > -L_D \ (0 < x_2 < R_D)\}$

It has been discussed in the beginning of the Section 3 and the detail derivation is given in Refs. [22,25]. We now consider the first step of effective transformation of shearing die flow to the elongational one at the cross-section $L^-$ located near the exit but still in the die. In this case, matching condition described by Eq.(13) is applicable. Applying Eq.(13) to match the different flow types at the die exit, we can find the values $\lambda_k^-$ $(k = 1, 2, ...)$ of the elastic tensors as:

$$x_1 = L^-: \quad \dfrac{L_W}{Q} \int_0^{L_D} v_1^{sh} (c_{11,k} + c_{22,k} + 1)^{n+1} dx_2 = [(\lambda_k^-)^2 + 2(\lambda_k^-)^{-1}]^{n+1}, \quad (k = 1, 2, ...); \tag{30}$$

$$x_1 = L^-: \quad \dfrac{2\pi}{Q} \int_0^{R_D} x_2 \cdot v_1^{sh} (c_{11,k} + c_{22,k} + 1)^{n+1} dx_2 = [(\lambda_k^-)^2 + 2(\lambda_k^-)^{-1}]^{n+1}, \quad (k = 1, 2, ...). \tag{30*}$$

Here the elastic Finger tensor profile ($c_{ij,k}^{sh}$) for the die shear flow at the exit is known from the die developing flow. It should be mentioned that the parameters

$\lambda_k^-$ $(k=1,2,...)$ found by this procedure, produce an effective non-zero elongation force $F^-$ at the cross section $x_1 = L^-$, even if the free extrudate flow happens when $F_0 = 0$. It means that the value $F^-$ describes the flow averaged longitudinal stress in the shearing flow. This is completely similar to the initial force at jet approach in the reservoir flow.

*3.3 Instant elastic recovery*.

As mentioned, an instant swell of extrudate happens just after the flow exits the die, and the swell further increases with the increase in traveling distance. In order to account these step variations, we calculation, we remind that $L^\pm$ are related to $x_1 = L \pm 0$, and corresponding values of parameters at these cross sections are:

$$x_1 = L^- : A^- = 2L_W \cdot L_D, \quad \text{or} \quad A = A^- = \pi R_D^2, \ \lambda_k = \lambda_k^- ; \ x_1 = L^+: \ A = A^+, \ \lambda_k = \lambda_k^+ \qquad (31)$$

In the cross-section $x_1 = L^-$ all the parameters are known; they are unknown in the cross section $x_1 = L^+$. To find these parameters we involve the instant recovery approach in the transition of polymer liquid between the cross sections $x_1 = L^\pm$. In this case, the following equations hold for both the geometries:

$$\lambda_k^+ = \lambda_k^- \lambda_s \ (k=1,2,...), \quad A^+ = A^- / \lambda_s, \quad A^- \sum_k G_k (I_{1k}/3)^n \cdot (\lambda_k^2 - 1/\lambda_k)\Big|_{\lambda_k = \lambda_k^- \lambda_s} = F_0 \lambda_s. \qquad (32)$$

The first relation in (32) is the definition of instant (elastic) recovery stretch $\lambda_s$, which is independent from the modal values of elastic stretches. The second relation in (32) is the relation for the step-wise variation for instant recovery, which also formally follows from Eq.(27) after integrating it for the step-variation to obtain $\lambda_k^- A^- = \lambda_k^+ A^+$, with the following use of the first relation in (32). The last relation in (32) simply reflects the fact that after instant recovery, the elongational force $F_0$ is given for the extended extrudate or absent for the free swelling extrudate at the cross section $x_1 = L^+$. The third relation in (32) serves for finding the step recovery stretch $\lambda_s$, the second relation for finding the initial cross section $A^+$ of the extrudate after instant recovery, and the first relation serves for finding initial value of elastic stretches $\lambda_k^+$.

As shown in (29), the values $\lambda_k^+$ and $A^+$ serve as initial condition for solving problem (27), (28).

*3.4. Numerical procedures*

They start with solving Eqs.(30) and (30*) for $\lambda_k^-$ using trapezoidal integration and the root finding subroutine. Another root finding subroutine was applied to solving the last equation in (32) to determine the value of instant elastic recovery stretch $\lambda_s$ and then find the values of $\lambda_k^+$ and $A^+$.

For solving Eqs. (27) and (28), they were numerically reformulated in explicit way using forward difference method and the unknown values of $\lambda_k$ and the area $A$ were found using root finding subroutine. Since the swelling ratio growing is critical at the initial stage, we used the composite grids scheme to axial direction assigning more dense meshes at the early stage. The growing length in specific direction is calculated from the calculations of extrudate area $A$ via Eqs(27) and (28).

## 4. Results and discussions

All calculations were made under isothermal condition with given flow rates. The calculations were exemplified using various polymers: polyisobutylene P20 [9], LLDPE [19] (FN1010 by ATOCHEM, $M_W$=120,000g/mol and polydispersity index equal to 6.3 at 160°C 0.76 g/cm³) for circular geometry and polyisobutylene Vistanex produced by Exxon [26] for the slit geometry. The Maxwell modes (discrete relaxation spectra) for LLDPE and Vistanex were taken from the literature data. It was confirmed that these modes could satisfactorily describe the experiments and/or direct calculations. Since even simple shear data for PIB 20 were only roughly theoretically described in [9], we assumed that the deviation of calculations from experiments in description of swelling data could result from incomplete characterization of relaxation spectra. Therefore instead of using original Maxwell modes of P20 from [9], we found them anew to fit better the steady simple shearing experimental data in [9], disregarding possible ill-posedness. We also tried to improve this procedure with mode refining.

Table 1 shows the values of numerical material parameter $n$ and specification of dimensionless function $b$ in CE's (1) and (2). It should mentioned that LDPE and LLDPE melts with long side chains show quite anomalous behavior in the extensional experiments possible due to the hardening effects [24], whereas they demonstrate a common behavior in the simple shearing. Moreover, LDPE/LLDPE melts exhibit much lower values of extensional viscosity in the repeated extension after relaxation of the first one. This might be attributed to mechano-degradation. Recently Ref. [27] reported that the predictions of LDPE/LLDPE shear and extensional flows were successful using the specification of the function $b$ shown in Table 1.

Table 2 shows the modified Maxwell modes for PIB-20 that we used in our calculations, as compared to the original modes (in bracket) found in [9]. Fig.1-A demonstrates the comparison of constitutive modeling with simple shear experimental data.

Figure 1-B compares our calculations and experiments [9] for free swelling development vs. the distance from the die. In computation of swelling for P-20, the exit flow is assumed to be developed because the original paper [9] did not provide necessary experimental geometry features. Based on the facts that the experiments in [9] were performed at L/2R$_D$ ratio 20 and the $De$ number ($= 4Q/\pi R_D^3 \cdot \overline{\theta}$) ranged 5~35, we expect our assumption of developed shear flow at the die exit to be close enough to the real experimental situations. We also should mention that the experiments were carried out under given pressure drop (wall shear stress). Therefore the flow rate was found from the flow curve (Fig. 1-A) for the given wall shear stress and then the swelling was computed using our scheme of calculations. One can see that with the use of redefined Maxwell modes, the results of calculation agree much better wit experiments. Also, calculated instant elastic recovery, which is not indicated in Ref. [9], is similar to Ref. [10].

Figure 2 presents our computational predictions of PIB Vistanex [26] free swelling for various die lengths and given flow rates. Unfortunately, there are no literature experimental data for making comparisons. As demonstration purpose to continue our previous paper, the computation results include for the swelling on the various die lengths. The detail experimental conditions and material characteristics are shown in Table 3, and the calculations in Fig.2 are performed for the die lengths having 0

(1), L/20 (2), L/10 (3), L/2 (4) and L (5) (where L=2.55cm). Since there is no significant elastic energy stored in the reservoir entrance region for low flow rates, the reservoir deformation effects dissipate quickly in the die entrance region. The first two plots in Fig. 2, which have $De$ numbers ($De = 3Q/2L_W L_D^2 \cdot \bar{\theta}$) equal to 8 for Fig2-A and 19 for Fig.2-B. Therefore except direct discharging case, the free swelling profiles are not well resolved, even if the flows exited die before they start to be developed. In faster flows, more elastic energy accumulates, so it results in more instant elastic recovery and in more swelling ratio. In these cases, the die length plays important role on the exit flow profile. When the die length is long enough, the flow is closed to be developed therefore keeps little memory of entrance deformation. On the contrary, when the die length is short enough, the die shear flow contain the memory of entrance effects that not dissipate with shortening die length. Therefore, the extrudate swell profiles after flow in short dies are well distinguished from that after developed flows in long dies. It is well indicated in last two plots in Fig. 2-C and 2-D, whose $De$ numbers are equal to 57 and 104 respectively. The results using very short dies (i.e. 0, L/20 and L/10) are found well separated from these calculated for relatively long dies (L/2 and L) for given flow rates. Obviously the calculated swelling profiles larger swelling ratio and longer developing distance at the higher $De$ number and for shorter dies.

There is another experimental data for high $De$ number free extrudate swelling after flow in short dies. These data were obtained in Ref.[19] for LLDPE. RMS 800 and elongation rheometer of rate controlled type were used in [19] for basic rheological measurements. Instron ICR3211 with electro-optical device (Zimmer OGH) was employed for swelling measurements. The details experimental setup for swelling test are presented in Table 4.

Figure 3 demonstrates the comparisons of CE (1) and (2) predictions with the basic rheological data. Using the dimensionless parameters and the Maxwell modes in Table 1 and 4, the simple shear (Fig.3-A), dynamic (Fig.3-B), elongation (Fig.3-C) and start-up (Fig.3-D) flow computations agree well with the experimental data.

Figure 4 shows the comparison the measured and calculated longitudinal velocity profiles in free swelling extrudate flow found from the cross-sectional area ratios,

$A^+/A(x_1)$. The swelling computations were performed using the data in Table 4. Instead of applying uncertain damping function that used to evaluate tapered entrance deformation with fitting parameters in Ref. [19], we used in our computation the quasi-1D model of contraction flow (the jet approach with secondary corner flow) neglecting shear deformation at tapered entrance. The computations in Fig. 4-A were performed with $De$ ($=4Q/\pi R_D^3 \cdot \bar{\theta}$) number about 40 at the flow rate $Q=$ 7.125e-10 m$^3$/s with L/2R$_D$=19.23 (Die 1 in Table 4). It is seen that the instant elastic recovery and the equilibrium cross-sectional area, obtained from the relation $A(x_1)/A_0 = v_1^0/v_1(x_1)$ at $x_1 \to L^+$ and $x_1 \to \infty$, respectively, are smaller than those in Fig. 4-B. This is because the computations for Fig.4-A showed that the die flow there is close to the developed one. On the contrary data in Fig. 4-B whose $De=$ 1300 at flow rate $Q$=2.375e-8 m$^3$/s for L/2R$_D$=0.96 (Die2 in Table4) showed that the die flow was not developed. Therefore in this case the calculations here were performed using our contraction/die flow model using the data in Table 4. Since the literature data did not provide the measurements in the die flow developing region, the possible tapered entrance effect on the developing flow could not been confirmed. This effect may generate additional shear deformation and affect the flow profiles. Nevertheless the both swelling experiments are successfully described with the jet-drag entrance model while all basic rheological data are well described with the chosen CE (1-2).

## 5. Conclusions

This paper analyzed the short die effect on swelling. Although there are plenty of published data for the die swelling of polymer fluids, they do not provide necessary information for comparison. Firstly the information of rheological characterization of the polymer melt needs to be provided which makes possible to evaluate the discrete relaxation spectrum $\{G_k, \theta_k\}$. Secondly the precise description of the geometry and measurements should be available. Additionally, polymer melts are more suitable for our modeling than polymer solutions. Although there is no problem to describe a polymer solution by a certain CE, it is hard to generate high enough $De$ number, when our 1-D contraction flow model works well. Even if generating high $De$ number flow is experimentally possible, the inertia effects of polymer solution (Reynolds number effects) are not negligible. Therefore we found from the literature only the presented here data that not enough to compare our model calculations to the direct calculations and/or experimental data.

Nevertheless, our 1-D contraction flow model which calculates the polymer flows in dies of arbitrary length taking into account highly nonlinear viscoelastic phenomena, allows us to consistently analyze the polymer flows in the whole flow region as one continuous process, involving no fitting parameters. We also demonstrated that our computationally low cost model works well at high enough *De* numbers, yet being applicable to other CE's. We also expect it to work well for the more complicated polymer flows, including non-symmetric flows, with relatively easy computations.

In our research, the isothermal fast complex flow with was considered. Even though the results seem to be satisfactory for the tested cases, there is still one more additional important factor that we need to take into account. This is the effect of non-isothermality with very high dissipative heat generation that plays very important role in high-speed (high *De* number) polymer processing operations. The extension of our quasi-1D flow model on analysis non-isothermal regimes of polymer flows will be reported elsewhere.

**Table 1.** Material Parameters in Eqs. (1) and (2)

| Polymer Melts | Parameters |
|---|---|
| Polyisobutylene (P-20) Leonov and Prokunin [9] | $n = 0.1$ $b = 1$ |
| Polyisobutylene (Vistanex) Isayev and Upadhyay [26] | $n = 0.1$ $b = 1$ |
| LLDPE (FN1010 by ATOCHEM) Fulchiron *et al* [19] | $n = 0.03$ $b = exp[-\beta\sqrt{(I-3)}] + \dfrac{sinh[\nu(I-3)]}{[\nu(I-3)+1]}$ $\beta = 0.15, \ \nu = 0.002$ |

**Table 2.** Material Parameters and Swelling Experimental Conditions for P20

| Polymer Melts | Parameters at $300^0K$ | | Conditions |
|---|---|---|---|
| | $\theta_k$ (sec) | $G_{k,}$ (KPa) | |
| Polyisobutylene P20 Leonov and Prokunin [9] | 140.741 (380) 3.208 (3.5) | 6.48 (2.70) 88.00 (78.00) | *Die Geometry* (cm) $2R_D = 0.5$ $L/R_D = 40$ |

**Table 3.** Material Parameters and Swelling Experimental Conditions for Vistanex.

| Polymer Melts | Parameters at $300^0K$ | | Geometry (cm) (Height x Depth x Width) |
|---|---|---|---|
| | $\theta_k$ (sec) | $G_k$, (KPa) | |
| Polyisobutylene (Vistanex) Isayev et al [26] | 7.025 1.553 0.182 | 4.55 11.59 86.81 | $L \times 2L_D \times L_W$ *Reservoir Dimensions* 4x0.822x2 *Die Dimensions* 2.55x0.121x2 |

**Table 4.** Material Parameters and Swelling Experimental Conditions for LLDPE (FN1010)

| Polymer Melts | Parameters at $323^0K$ | | Geometry (cm) |
|---|---|---|---|
| | $\theta_k$ (sec) | $G_k$, (KPa) | |
| LLDPE (FN1010 by ATOCHEM) Fulchiron et al [19] | 1.28E-4 6.12E-3 4.10E-2 2.77E-1 2.01 1.57E1 1.35E2 | 1849.00 219.90 82.02 16.94 1.85 0.13 0.0071 | Reservoir Radius ($R_R$) = 0.476 Die Radius ($R_D$) = 0.0625 Die 1: Die Length ($L$) = 0.12 Die 2: Die Length ($L$) = 2.4 |

**Figure Captions**

**Fig. 1** PIB (P20) Experimental data at $300^0$K and calculation with modified Maxwell modes shown in Table 2: **A** - simple shear experiments, **B** - die swelling profiles at various flow rates.

**Fig. 2** Calculations of free die swell profiles for PIB (Vistanex) with given flow rates $Q$ equal to: **A** - 1.0e-8 m$^3$/s, **B** - 2.4e-8 m$^3$/s, **C** - 7.1e-8 m$^3$/s, **D** - 1.3e-7 m$^3$/s . The lines stand for for different die lengths: **1** - 0, **2** - L/20, **3** - L/10, **4** - L/2, and **5** - L (L=2.55cm) indicated in Table 3.

**Fig. 3** Basic rheological experimental data for LLDPE (FN1010) at 433K : **A** - steady simple shearing, **B** - dynamic test, **C** - uniaxial extension, **D** - startup shear flow at shear rate 1s$^{-1}$. Symbols-experiments and lines-caculations.

**Fig. 4** Axial velocity along flow direction for LLDPE melt under various die swelling conditions at $433^0$K : **A** - flow rate $Q = 7.125$e-10 m$^3$/s with $L/2R_D = 19.23$, **B** - flow rate $Q = 2.375$e-8 m$^3$/s with $L/2R_D = 0.96$.

Fig.1

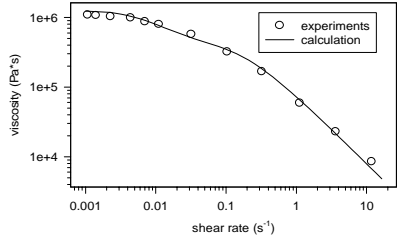

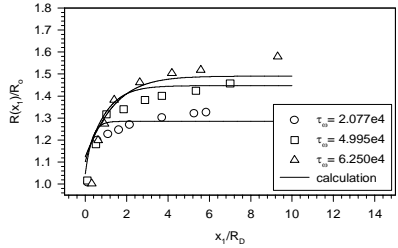

Fig.2

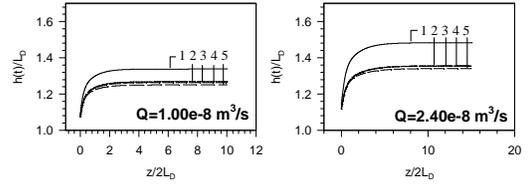

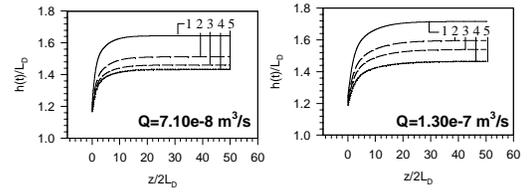

Fig.3

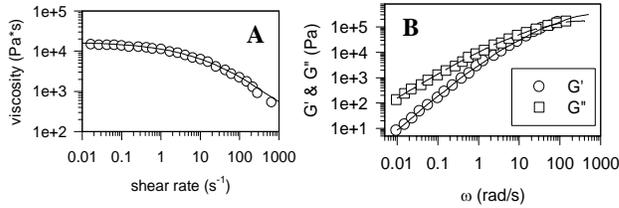

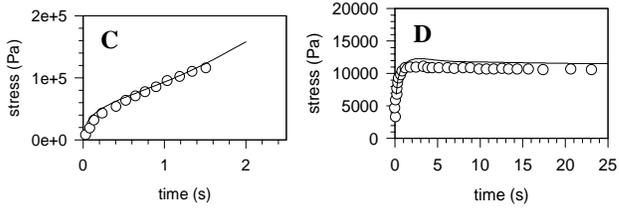

Fig.4

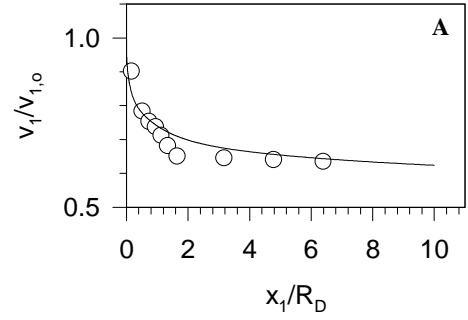

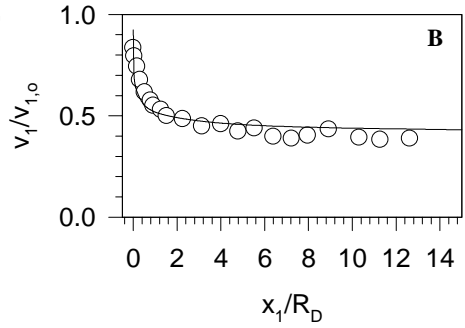